\input harvmac
\noblackbox

\input epsf

\def\tilde{\widetilde}
\newcount\figno
\figno=0
\def\fig#1#2#3{
\par\begingroup\parindent=0pt\leftskip=1cm\rightskip=1cm\parindent=0pt
\baselineskip=11pt
\global\advance\figno by 1
\midinsert
\epsfxsize=#3
\centerline{\epsfbox{#2}}
\vskip 12pt
{\bf Fig.\ \the\figno: } #1\par
\endinsert\endgroup\par
}
\def\figlabel#1{\xdef#1{\the\figno}}
\def\encadremath#1{\vbox{\hrule\hbox{\vrule\kern8pt\vbox{\kern8pt
\hbox{$\displaystyle #1$}\kern8pt}
\kern8pt\vrule}\hrule}}

\def\Im{{\rm Im}}

\def\pl#1#2#3{Phys. Lett. {\bf B#1} (#2) #3}


\font\cmss=cmss10
\font\cmsss=cmss10 at 7pt
\def\rlx{\relax\leavevmode}
\def\inbar{\vrule height1.5ex width.4pt depth0pt}
\def\IC{\relax\,\hbox{$\inbar\kern-.3em{\rm C}$}}
\def\IN{\relax{\rm I\kern-.18em N}}
\def\IP{\relax{\rm I\kern-.18em P}}
\def\ZZ{\rlx\leavevmode\ifmmode\mathchoice{\hbox{\cmss Z\kern-.4em Z}}
 {\hbox{\cmss Z\kern-.4em Z}}{\lower.9pt\hbox{\cmsss Z\kern-.36em Z}}
 {\lower1.2pt\hbox{\cmsss Z\kern-.36em Z}}\else{\cmss Z\kern-.4em
 Z}\fi} 
\def\IZ{\relax\ifmmode\mathchoice
{\hbox{\cmss Z\kern-.4em Z}}{\hbox{\cmss Z\kern-.4em Z}}
{\lower.9pt\hbox{\cmsss Z\kern-.4em Z}}
{\lower1.2pt\hbox{\cmsss Z\kern-.4em Z}}\else{\cmss Z\kern-.4em
Z}\fi}

\def\narrowplus{\kern -.04truein + \kern -.03truein}
\def\narrowminus{- \kern -.04truein}
\def\narrowminussub{\kern -.02truein - \kern -.01truein}

\def\half{{1\over 2}}

\def\cl{\centerline}

\def\O{{\Omega}}

\def\b{{\beta}}
\def\k{{\kappa}}
\def\a{{\alpha}}
\def\g{{\gamma}}
\def\m{{\mu}}

\def\d{{\delta}}
\def\o{{\omega}}
\def\G{{\Gamma}}
\def\ph{{\phi}}
\def\t{{\theta}}
\def\l{{\lambda}}

\def\Ph{{\Phi}}
\def\rh{{\rho}}
\def\ch{{\chi}}
\def\c{{\chi}}
\def\s{{\sigma}}

\def\r{{\rightarrow}}

\def\frac#1#2{{#1\over #2}}

\def\CD{{\cal D}}

\def\CF{{\cal F}}

\def\IZ{\relax\ifmmode\mathchoice
{\hbox{\cmss Z\kern-.4em Z}}{\hbox{\cmss Z\kern-.4em Z}}
{\lower.9pt\hbox{\cmsss Z\kern-.4em Z}}
{\lower1.2pt\hbox{\cmsss Z\kern-.4em Z}}\else{\cmss Z\kern-.4em
Z}\fi}
\def\IB{\relax{\rm I\kern-.18em B}}
\def\IC{{\relax\hbox{$\inbar\kern-.3em{\rm C}$}}}
\def\ID{\relax{\rm I\kern-.18em D}}
\def\IE{\relax{\rm I\kern-.18em E}}
\def\IF{\relax{\rm I\kern-.18em F}}
\def\IG{\relax\hbox{$\inbar\kern-.3em{\rm G}$}}
\def\IGa{\relax\hbox{${\rm I}\kern-.18em\Gamma$}}
\def\IH{\relax{\rm I\kern-.18em H}}
\def\II{\relax{\rm I\kern-.18em I}}
\def\IK{\relax{\rm I\kern-.18em K}}
\def\IP{\relax{\rm I\kern-.18em P}}

\def\p{\partial}

\font\cmss=cmss10 \font\cmsss=cmss10 at 7pt
\def\IR{\relax{\rm I\kern-.18em R}}

%

%
%
\def\eqnn#1{\xdef #1{(\secsym\the\meqno)}\writedef{#1\leftbracket#1}%
\global\advance\meqno by1\wrlabeL#1}
\def\eqna#1{\xdef #1##1{\hbox{$(\secsym\the\meqno##1)$}}
\writedef{#1\numbersign1\leftbracket#1{\numbersign1}}%
\global\advance\meqno by1\wrlabeL{#1$\{\}$}}
\def\eqn#1#2{\xdef #1{(\secsym\the\meqno)}\writedef{#1\leftbracket#1}%
\global\advance\meqno by1$$#2\eqno#1\eqlabeL#1$$}

\lref\rseiberga{N. Seiberg, ``New Theories in Six Dimensions and 
Matrix Description of M-theory on 
$T^5$ and $T^5/Z_2$,''
\pl{408}{1997}{98}, hep-th/9705221.}
 
\lref\rseibergb{O. Aharony, M. Berkooz, D. Kutasov and  N. Seiberg,
``Linear Dilatons, NS5-branes and Holography,''  
JHEP {\bf10} (1998) 004, hep-th/9808149. }

\lref\rseibergc{ O. Aharony, M. Berkooz, S. Kachru, N. Seiberg, 
E. Silverstein, ``Matrix Description of Interacting Theories in 
Six Dimensions,'' Adv.Theor.Math.Phys. {\bf 1} (1998) 148, 
hep-th/9707079.}

\lref\rwittenb{ E. Witten, ``On The Conformal Field Theory Of The 
Higgs Branch,'', JHEP {\bf 07} (1997) 003, hep-th/9707093.}

\lref\rms{J. Maldacena and A. Strominger, 
``Semiclassical decay of near extremal fivebranes,'' 
JHEP {\bf12} (1997) 008, hep-th/9710014.} 

\lref\rpp{A. Peet and J. Polchinski, ``UV/IR Relations in AdS Dynamics,''
Phys.Rev. {\bf D59} (1999) 65006, hep-th/9809022.}

\lref\rklebc{S. Gubser, I. Klebanov, A. Tseytlin, 
``String Theory and Classical Absorption by Threebranes,''
Nucl.Phys. {\bf B499}
 (1997) 217, hep-th/9703040.}

\lref\rklebd{S. Gubser and  I. Klebanov, 
``Absorption by branes and Schwinger
terms in the World Volume Theory,'' Phys.Lett. 
{\bf B413} (1997) 41, hep-th/9708005. }

\lref\rmalda{J. Maldacena, ``The large N limit of Superconformal 
theories and
Supergravity,'' Adv.Theor.Math.Phys. {\bf2} (1998) 231, hep-th/9711200.}

\lref\rcallan{C. Callan, J. Harvey and A. Strominger, ``Supersymmetric
String Solitons,''
hep-th/9112030, Lectures at the 1991 Trieste Spring School on 
String Theory and Quantum Gravity.} 
\lref\rtownsend{G. Gibbons, G.  Horowitz and P. Townsend,
``Higher-dimensional resolution of dilatonic black hole singularities,''
Class.Quant.Grav. {\bf12} (1995) 297, hep-th/9410073.} 

\lref\rwitten{E. Witten, ``Anti De Sitter Space And Holography,'' 
Adv.Theor.Math.Phys. {\bf2} (1998) 253, hep-th/9702150.}

\lref\rpol{ S. Gubser, I. Klebanov and A. Polyakov, 
`` Gauge theory Correlators from Non-Critical String Theory,''
 Phys.Lett. {\bf B428} (1998) 105, hep-th/9802109.}

\lref\rbanks{O. Aharony and T. Banks, ``Note on the Quantum Mechanics
of M Theory,'' JHEP {\bf 03} (1999) 016, hep-th/9812237.} 

\Title{ \vbox{\baselineskip12pt\hbox{hep-th/9904142}
\hbox{PUPT-1859}
\hbox{IAS-SNS-HEP-99-37}}}
{\vbox{\centerline{Comments on the IIA NS5-brane}
}}

\centerline{Shiraz Minwalla\footnote{$^1$}{minwalla@princeton.edu}
}
\smallskip
\centerline{\sl Department of Physics, Princeton University}
\centerline{\sl Princeton, NJ 08544, USA}
\medskip
\centerline{and}
\medskip
\centerline{Nathan Seiberg\footnote{$^2$}{seiberg@sns.ias.edu}}
\smallskip
\centerline{\sl School of Natural Sciences, Institute for Advanced Study}
\centerline{\sl Olden Lane, Princeton, NJ 08540, USA}

\vskip 0.8cm

\centerline{\bf Abstract}
\medskip
\noindent 
We study $N$ coincident IIA NS5-branes at large $N$ using
supergravity. We show that the absorption cross section for gravitons
in this background does not vanish at zero string coupling for
energies larger than ${m_s\over \sqrt{N}}$ ($m_s$ is the string
scale). Using a holographic description of the intrinsic theory of the
IIA NS5-branes, we find an expression for the two point function of the
stress energy tensor, and comment on its structure.

\vskip 0.5cm
\Date{April 1999}

\newsec{Introduction}

In this note we study $N$ coincident NS5-branes \rcallan\ in type II
theory. In the limit $g_0 \r 0$, $m_s$ fixed (where $g_0$ is the
asymptotic value of the string coupling and $m_s$ is the string mass)
the theory is free in the bulk. However, modes living on the 5-brane
continue to interact amongst themselves, while decoupling from the
bulk, defining a mysterious non-gravitational six-dimensional theory
\rseiberga\ (this construction was motivated by
\ref\brs{M.~Berkooz, M.~Rozali and N.~Seiberg, ``Matrix description of
M theory on $T^4$ and $T^5$,'' Phys. Lett. {\bf B408} (1997) 105,
hep-th/9704089.}), sometimes referred to as the little string theory.
Upon compactification, this theory inherits the $T$ duality of type II
string theory and is therefore nonlocal at the scale $m_s$.

Maldacena and Strominger \rms\ studied the system of $N$
non-extremal coincident NS5-branes with excitation energy density $\m
m_s^6$.  The classical solution for this configuration possesses an
asymptotically flat region, which turns into a tube on approaching the
5-brane. On descending down the tube one encounters the horizon of the
black branes. The local string coupling outside the horizon is
everywhere less than $\sqrt{{N\over \m}}$, and all curvatures in this
region are less than ${m_s\over \sqrt{N}}$. Therefore, when $\m\gg N
\gg 1$ semi-classical gravity is accurate outside the horizon.  In
particular, the brane Hawking radiates at temperature ${1\over
2\pi}{m_s\over \sqrt{N}}$, and modes behind the horizon continue to
couple to modes in the tube even for $g_0=0$.

This fact was interpreted in \rseibergb\ as a manifestation of
holography in the underlying string theory.  In particular, the
decoupled theory of the NS5-branes is the holographic projection,
along the lines of \rmalda, of the near horizon geometry
($g_0\rightarrow 0$ limit) including the long tube\foot{Similar ideas
have been suggested by various people, including C.V. Johnson,
J. Maldacena and A. Strominger.}.  It was further proposed in
\rseibergb\ that some of the observables of the NS5-brane theory are
the on-shell particles (vertex operators in the string theory limit)
in the bulk.  Unlike the situation in $AdS$, our geometry admits the
definition of an $S$ matrix.  Hence, off-shell six-dimensional
``Green's functions'' of these observables are identified as the $S$
matrix elements of the corresponding higher dimensional bulk
particles.

It should be emphasized that the boundary of the near horizon geometry
of the Euclidean NS5-brane is $R^6\times S^3$, and reduces to $R^6$
only after a Kaluza Klein reduction on the $S^3$. This is in contrast
with the near horizon geometry of, say, the Euclidean M5-brane, whose
boundary is $S^6$ rather than $S^6\times S^4$ (because the ratio of
the size of the $S^4$ to that of the $S^6$ goes to zero on approaching
the boundary).

The Euclidean near horizon geometry of the NS5-brane consists of a
semi-infinite tube, capped at the bottom by (Euclidean) $AdS_7 \times
S^4$. Let the analytic continuation of this space to Lorentzian
signature be denoted by $E$. $E$ possesses a future horizon $H^+$ and
a past horizon $H^{-}$ each at finite affine distance, and so is
geodesically incomplete. Its metric may be completed (see Appendix D)
by gluing together a new copy of $E$ at each of the two original
horizons, and then continuing this procedure indefinitely (just as the
AdS cylinder may be constructed by gluing together copies of Poincare
patches at their horizons). The resulting Penrose diagram is shown in
fig.\ 1, and contains an infinite number of pairs of $\CI^{\pm}_i$s,
differentiated in our notation by the subscript $i$. $\CI^{\pm}_i$,
together with the horizons $H^{\pm}_i$ constitute the boundary of the
$i^{th}$ wedge $E_i$.  It is not clear to us whether this analytic
continuation is physically relevant\foot{We thank O. Aharony and
T. Banks for a useful discussion on this point.}.

\fig{The Penrose diagram of the geodesic completion of the metric of
the NS5-brane.} {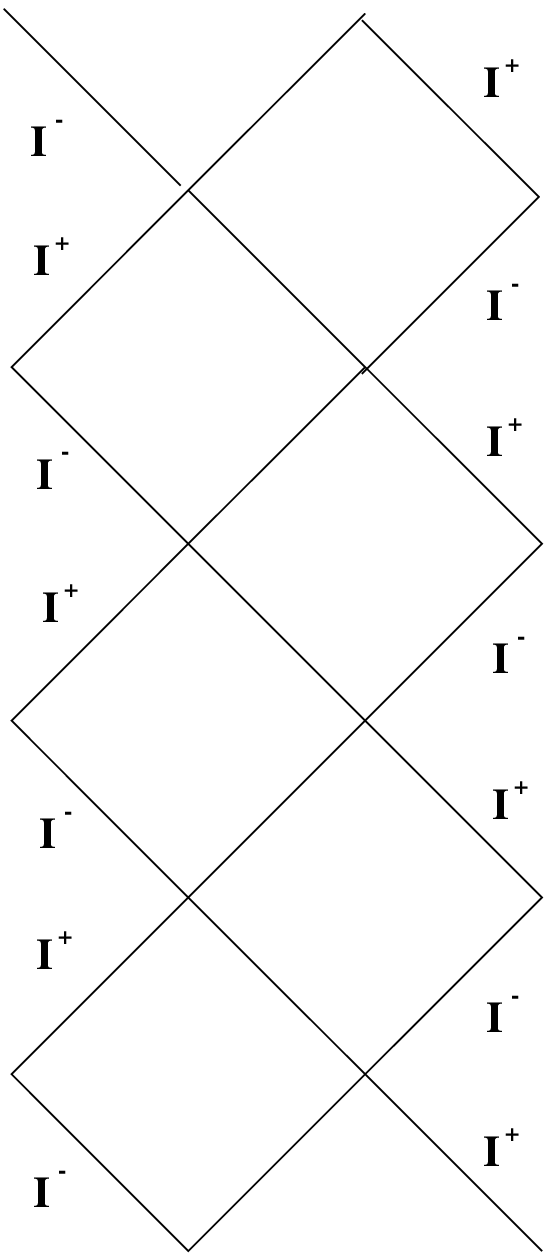}{1.4truein}

The correlation functions defined in \rseibergb\ in terms of $S$
matrix elements are naturally found in momentum space.  However, it
was pointed out by Peet and Polchinski \rpp\ that these answers cannot
be Fourier transformed to coordinate space, and therefore the little
string theory is nonlocal.  The reason for that is that the momentum
space answers are obtained after a momentum dependent multiplicative
renormalization with the characteristic scale $m_s\over \sqrt N$.  We
will return to this below.

Aharony and Banks \rbanks\ computed the entropy of the little string
theory as a function of energy $\omega$ using its DLCQ definition
\refs{ \rseibergc, \rwittenb} and found $6\sqrt{N}\omega \over m_s$, in
agreement with the Bekenstein-Hawking entropy computed in \rms. This
entropy formula suggests that the number of states in the system grows
rapidly with energy, thus explaining why the typical momentum space
correlation function cannot be Fourier transformed \rpp.  It also
reproduces the scale of nonlocality as $m_s\over \sqrt{N}$.

In this paper we study the propagation of a mode of the graviton,
which propagates like a minimal scalar in the supergravity background
of $N$ coincident IIA NS5-branes. Its $S$ matrix elements are
interpreted as the correlation functions of the energy momentum tensor
of the little string theory.  In section 2 we discuss the validity of
the supergravity approximation, define variables, and set up our
equations. In section 3 we compute the absorption cross section of a
graviton incident onto the 5-branes from the tube, and note that it is
nonzero at $g_0=0$ for energies larger than ${m_s\over \sqrt{N}}$.
This result strengthens that of \rms. In section 4 we follow
\rseibergb\ to derive an expression for the two point function of the
energy momentum tensor of the little string theory, and comment on its
structure. In section 5 we explain the relation between our results
and previous calculations.  In Appendix A we estimate the high energy
behavior of the two point function of section 4 using the WKB
approximation. In Appendix B we develop a low energy expansion for the
two point function. In Appendix C we study a simple toy model, similar
in some ways to the NS5-brane.  In Appendix D we discuss the causal
structure of the geodesic completion of the near horizon geometry of
the NS5-brane.

\newsec{The Setup}

\subsec{The Background Metric}   

The classical string frame background corresponding to $N$ coincident
extremal NS5-branes is
\eqn\etrica{\eqalign {ds^2&=dx_6^2+(1+{N\over m_s^2r^2})
(dr^2 +r^2d\O_3^2) \cr
e^{2\Ph}&=g_0^2(1+{N\over m_s^2r^2}).\cr }}
This background possesses an asymptotically flat region $r\gg
{\sqrt{N}  \over m_s}$
connected to a semi-infinite flat tube $r\ll {\sqrt{N}\over m_s} $, 
with the topology of $R^+\times S^3\times R^6$ (these factors are
parametrized by $r$, $\Omega$ and $x_6$ respectively).

In this paper we will restrict our attention to the IIA version of the
little string theory.  Its moduli space of vacua is ${(R^4\times
S^1)^N\over S_N}$ corresponding in M theory to the positions of the
$N$ M5-branes in the transverse space and on the $11^{th}$ circle.
\etrica\ is the solution that corresponds to stacking all branes at 
the origin in the $R^4$, but smearing them evenly over the $S^1$,
and therefore does not correspond to  any true vacuum 
(single point in moduli space) of the theory. 

We wish to study the theory at the most singular point in its moduli
space, where all the branes are on top of each other both in $R^4$ and
$S^1$.  The corresponding classical background is most conveniently
written in eleven dimensions
\ref\sunny{N. Itzhaki, J.M. Maldacena, J. Sonnenschein and 
S. Yankielowicz, ``Supergravity and the Large $N$ Limit of Theories
with Sixteen Supercharges,'' Phys.Rev. {\bf D48} (1998)
46, hep-th/9802042.}
\eqn\metricg{ds^2=A^{-{1\over 3}}dx_6^2+A^{{2\over 3}}(dx_{11}^2+dr^2+
r^2d\O_3^2)}
\eqn\a{A=1+\sum_{n=-\infty }^{\infty}{ N\pi l_p^3 \over
[r^2+(x_{11}-2\pi nR_{11})^2]^{{3\over 2}} }.}  The $11^{th}$
dimension is compact: $x_{11}=x_{11}+2\pi R_{11}$ and $l_p$ is the
eleven-dimensional Planck length.  For $r \gg R_{11}$ the summation
above may be replaced by an integral and we recover \etrica, on
dimensional reduction and transforming to the string frame. For $r\ll
R_{11}$ a single term in the summation in \a\ dominates, and \metricg\
reduces to the near horizon of $N$ stacked M5-branes.  Therefore
\metricg\ represents asymptotic flat space connected to $AdS_7 \times
S^4$ by a long tube.

\subsec{ Validity of Supergravity} 

We start by considering the general case of M theory compactified on
a circle, whose physical radius varies in an arbitrary spatially
dependent fashion in the ten non-compact directions. We then
specialize to M theory on \metricg.

When the physical length $r_{11}$ of the M theory circle is much
larger than $l_p$ (the eleven-dimensional region) eleven-dimensional
supergravity is valid for energies and curvatures much smaller than
${1\over l_p}$.  When $r_{11}\ll l_p$ (the ten-dimensional region) the
local IIA string coupling $g=({r_{11} \over l_p})^{{3\over 2}}$ is
small, and string perturbation theory is valid.  It reduces to IIA
supergravity for energies and curvatures much smaller than the string
scale. Under these conditions proper energies are smaller than
$m_{eff}={g\over r_{11}}\ll {1\over r_{11}}$.  Therefore, no Kaluza
Klein modes on the $11^{th}$ circle are excited, and IIA supergravity
is identical to eleven-dimensional supergravity.

In summary, eleven-dimensional supergravity may be used over all space
for energies $\o \ll { 1\over l_p}, m_s$ provided curvatures are small
in string units in the ten-dimensional region, and in Planck units in
the eleven-dimensional region.

Throughout this paper $g_0$ is taken to be very small, 
so that $R_{11}\ll l_p \ll {1\over m_s}$ and the  energy condition above
is satisfied if $\o \ll m_s$. 
In the ten-dimensional region in \metricg, $r\gg R_{11} \sqrt{N}$,
the curvatures in string units are always less than ${1\over \sqrt{N}}$. 
Curvatures in the eleven-dimensional region $r\ll R_{11} \sqrt{N}$
are always less than $N^{-{1\over 3}}$ in Planck units. 

Hence 11d supergravity is valid in \metricg\ for $\o \ll m_s$
and $N \gg 1$. Therefore, we study the theory in the 't Hooft limit 
$m_s \r \infty$, $N \r \infty$ with 
${m_s \over \sqrt{N}}$ fixed\foot{This is the usual 't Hooft limit 
for the low energy gauge theory of the IIB little string theory.}. 
In this limit classical supergravity is valid at all energies. 

\subsec{ Qualitative motion}

We focus initially on the tube region, $R_{11}\ll r \ll {\sqrt{N}
\over m_s}$, where the expression for $A$ \a\ simplifies to $A\approx
{N\over m_s^2r^2}$.  A mode of energy $\o$ at asymptotic infinity has
proper energy $\o A^{{1\over 6}}$. The local proper radius of the
$11^{th}$ circle is $r_{11}=R_{11}A^{{1\over 3}}$. The local value of
the string coupling $g$ is $g_0 A^{{\half}}$. The local effective
string tension (proper energy per unit proper length) is given by
$m_{eff}^2=m_s^2A^{{1\over 3}}$.

The proper energy of a Kaluza Klein excitation in the $11^{th}$
direction is ${1\over r_{11}}={1\over R_{11}A^{1\over 3}}$. This
corresponds to energy at infinity $\o={1\over R_{11}A^{{1\over
2}}}={r\over R_{11}}{m_s\over \sqrt{N}}$.  Therefore, Kaluza Klein
modes are excited at the bottom of the tube for energies larger than
$m_s\over \sqrt N$.

A graviton with polarization parallel to the brane propagating in
\metricg\ with no momentum along the $S^3$ or the brane directions
obeys at the quadratic level the equation of motion of a minimally
coupled scalar \rklebc.  Such a graviton couples to a particular
polarization of the stress energy tensor on the brane world
volume\foot{Upon compactification the theory on the brane has more
than one ``stress tensor'' \refs{\rseiberga,\rseibergb} and this is
one of them.}.  We denote it by $\phi$ and study its propagation in
\metricg.  In the tube 
(${\sqrt{N} \over m_s}\gg r \gg R_{11}$) we change coordinates to
$z={\sqrt{N}\over m_s}\ln({rm_s\over \sqrt{N}})$ and ignore variations 
in the $x_{11}$ direction to find the string frame metric and coupling
constant
\eqn\metricb{ds^2=dx_6^2+dz^2 + {N\over m_s^2}d\O_3^2}
\eqn\dilatonb{g=g_0e^{-{z m_s\over \sqrt{N}}}.}
The Einstein frame metric is 
\eqn\metricf{ds^2=e^{{z m_s \over 2\sqrt{N}}}(dx_6^2+dz^2 + {N
\over m_s^2}d\O_3^2),}  
and the quadratic action for $\phi$ is
\eqn\actt{S={(2\pi)^5\over \k_{10}^2}
\int d^{10}x e^{{2z m_s \over \sqrt{N}}}[(\p_0\phi)^2
-(\p_{z}\phi)^2],}
where we have chosen a convenient normalization. 
In terms of $\tilde \phi=e^{{z m_s \over \sqrt{N}}}\phi$ it is
\eqn\acttt{S={(2\pi)^5 \over \k_{10}^2}\int d^{10}x [(\p_0 \tilde
\phi)^2 -(D_{z}\tilde\phi)^2],}
where $D_{z}=\p_{z}-{m_s\over \sqrt{N}}$.
This action leads to the equation of motion 
\eqn\new{(-\p_0^2+\p_z^2-{m_s^2 \over N})\tilde \ph=0,}
corresponding to a free massive particle with mass ${m_s \over
\sqrt{N}}$.  The two independent solutions in the tube are
$\phi_\pm=e^{{m_s\over \sqrt N }(\b_{\pm}(s)z -i\sqrt s t)}$, with
$$\b_{\pm}(s)=-1 \pm \sqrt{1-s}$$
where
$$s={\o^2 N \over m_s^2}$$ 
is energy squared in units of the mass gap.
 
The string frame length of the tube is 
${\sqrt{N} \over m_s} \ln({\sqrt{N}
\over g_0})$, and goes to infinity as $g_0 \r 0$. 
We set $g_0=0$ and then the asymptotic region is driven off to 
infinity, and can be ignored. Our geometry consists of an infinite tube 
capped at the bottom by $AdS_7 \times S^4$. 

A particle with $s<1$ cannot propagate in the tube and is confined 
in the AdS part of our geometry. A particle with $s>1$ can 
propagate in the tube, and can leak out of the AdS region into the tube. 
Consequently, absorption from the tube into the AdS region is 
also possible.

\subsec{ The Differential Equation}

When studying the near horizon region $r\ll {\sqrt{N} \over m_s}$ 
it is convenient to rescale the coordinates, 
$\rh={r\over R_{11}}$, $\c ={x\over R_{11}}$. 
$y_i={x_im_s\over \sqrt{N}}$, where $i$ runs over the 5+1 dimensions parallel 
to the brane. \metricg\ becomes
\eqn\metrici{ds^2=(N\tilde{A})^{{2\over 3}}l_p^2
[ \tilde{A}^{-1}dy_6^2+
d\ch^2+d\rh^2+\rh^2d\O_3^2 ]}  
$$\tilde{A}=\sum_{n=-\infty }^{\infty}{ 
\pi\over [\rh^2+(\ch-2\pi n)^2]^{{3\over 2}}}.$$
$\c$ has periodicity $2\pi$. The tube is the region $\rh \gg 1$, while 
$\rh \ll 1$ is the M5 part of the geometry.

The propagation of a complex minimally coupled scalar is 
governed by the action 
\eqn\actmin{S_b={(2\pi)^5\over \k_{11}^2}\int d^{11}x \sqrt{g}
|\del \ph|^2.}

We set $\ph=\t e^{-i\o t}=\t e^{-i\sqrt{s}y_0}$ with $\t$ constant 
on the 3-sphere and the spatial directions parallel to the 5-brane.  
\actmin\ becomes 
\eqn\rescact{S_b={m_s^6V\over 2\pi}\int d\ch d\rh \rh^3 (
-\tilde{A}s|\t|^2+|\p_{\ch}\t|^2 +|\p_{\rh}\t|^2 ),}where $V=\int d^6x
={N^3\over m_s^6}\int d^6y$ 
is the spacetime volume of the brane. 
$\t$ obeys the equation of motion
\eqn\mmeqgc{\p_{\ch}^{2}\t+{1\over {\rh}^{3}}\p_{{\rh}}{\rh}^{3}
\p_{{\rh}}\t+
\sum_{n=-\infty }^{\infty}{s \pi \over [{\rh}^{2}+(\ch-2\pi n)^2
]^{{3\over 2}}}\t=0,}
which is valid for all $\rh$ and  
on the full complex $s$ plane.  
$s$ is real and positive in Minkowski
space, and real and negative in Euclidean space. Recall that the 
mass gap in the tube is at $s=1$.

\subsec{A particular solution}

Define $f(s, \rh, \ch)$ as the unique solution to \mmeqgc\ obeying:

\item{1.} Near $\rh, \ch=0$ $f=f(s,u)$, where $u^2=\rh^2+\ch^2$. 
Further $f(s,u=0)=0$ in Euclidean space. 
This condition ensures that in Minkowski space at small $u$, 
$f$ is a wave that carries flux only into the 5-branes.
It uniquely determines $f(s, \rh, \ch)$ up to a 
normalization everywhere on 
the complex $s$ plane. In particular at small $u$ 
\eqn\form{f\approx C(s)\left({-s\pi\over u}\right)^{3\over 2}K_3
\left( 2 \sqrt{{-s\pi \over u}}\right)} 
\item {2.} $f$ is normalized such that for $\rh \gg 1$
\eqn\fgform{f(s, \rh, \ch)\approx \rh^{\b_+(s)}+ D(s)\rh^{\b_{-}(s)}.}

$C(s), D(s)$ are determined in principle by \mmeqgc. 
In the absence of 
a complete solution we list what we know about these two functions. Note
that $C(s)$ and $D(s)$ are real on the negative real axis.

\item{a.} For real $s$ $f(s, \rh,\ch)$ obeys a flux conservation equation
derived from \mmeqgc. The equation is trivial for negative 
$s$, but nontrivial for positive $s$ yielding
\eqn\constr{\Im\big[ (1+D(s))(1-D^*(s))\sqrt{1-s}\big] =
{ \pi^3 s^3 |C(s)|^2  \over 6}.}

\item{b.} In Appendix A we use the WKB approximation to argue 
that on the positive real axis 
$$\lim_{s\r \infty}|D(s)| =0.$$

\item{c.} At $s=0$ \mmeqgc\ becomes an equation for free propagation of 
waves in 4 spatial dimensions and may be exactly solved.  $C(s)$ and
$D(s)$ may then be determined for small $s$ by perturbing around this
exact solution.  In Appendix B we show that in perturbation theory
$C(s)$ and $D(s)$ take the form (a similar result was anticipated in
\rseibergb)
\eqn\cform{C(s)=\sum_{n=0}^{\infty}(s^3\ln s)^n f_n(s)}
\eqn\dform{D(s)=\sum_{n=0}^{\infty}(s^3\ln s)^n g_n(s),}
where $f_n(s)$ and $g_n(s)$ are analytic functions at $s=0$ 
related to each other by \constr. 
Explicitly computing the first few terms in the perturbative expansion 
(Appendix B) we find   
$$f_0(s)=1+\half [\g -\ln (4\pi)]s +\CO(s^2)$$
$$g_0(s)=\CO(s^2)$$
$$f_1(s)={\zeta(3) \over 48 \pi}+\CO(s)$$
$$g_1(s)={\pi^2 \over 12}+\CO(s)$$
where $\g$ is Euler's constant.  Therefore, 
$$C(s)=1+smaller ~terms,$$
$$D(s)= Analytic +{\pi^2 \over 12} {s^3\ln s } + smaller ~terms.$$
To leading order $\Im D(s)={\pi^3 \over 12}s^3|C(s)|^2$ in accordance
with \constr.

\newsec{Absorption from the tube for $s>1$}

When $s>1$ particles propagate in the tube. 
$f(s, \rh, \ch)$ for positive $s$ is a solution with unit flux incident 
down into the tube, flux $|D(s)|^2$ is  
reflected back up the tube, and flux $1-|D(s)|^2$ is absorbed 
into the M5-branes. Therefore, if a particle moves down the 
tube, the probability that it will be reflected back up the tube  is 
$|D(s)|^2$. The reflection $S$ matrix element is $2\sqrt{s-1}D(s)$ 
(see Appendix  C for our normalization of  the $S$ matrix). 

Since particles with $s>1$ can be absorbed into the M5-branes, the 
reverse process -- the leakage of particles with $s>1$ out of the AdS region
into the tube -- is also possible. This is unlike the situation 
with a stack of  M5-branes
in flat space, which ceases to absorb and emit particles in the decoupling 
limit $l_p \r 0$. 

This observation strengthens the Maldacena-Strominger \rms\
non-decoupling from the tube in two ways. First, the non-decoupling
occurs at finite energy and does not require energy densities. Second,
the non-decoupling occurs at energies above ${m_s \over \sqrt{N}}$.
This value is below ${m_s \over N^{1\over 6}}$, which is required for
the validity of the Maldacena-Strominger approximations.

\newsec{Two Point Functions} 

\subsec{The Prescription}

In this section we will use the holographic proposal of 
\rseibergb\ to find an expression for the two point function of the 
little string theory operator $O$ that couples to our minimal scalar
according to 
\eqn\aaa{ S_{int}=\int d^{6}x (\t^*(x,L) O(x)+\t(x,L) O(x)^*).}  
Since $\t$ is a mode of the graviton, we interpret $O$ as a component
of the energy momentum tensor of the brane.  We will be working in
momentum space and use
$$O(k)=\int {d^{6}x\over (2\pi)^{3}} O(x)e^{ikx}.$$ 
Let $\langle O(k)O^*(-k') \rangle =\Pi(k)\d(k-k')$, and define a
dimensionless two point 
function $\Pi^L(s)={1\over m_s^6} {\Pi(k)}$ at ${k^2N \over m_s}=s$.

$h(s,\rh,\ch)={f(s,\rh,\ch) \over f(s, L, \ch=0)}$ (for $L\gg 1$ the
$\chi$ dependence of $f$ is exponentially small) is a
solution of
\mmeqgc, which is regular everywhere in Euclidean space, and is unity
at $\rh=L$.  According to the prescription of \refs{\rpol, \rwitten}
adapted to our situation, $\Pi^L$ may be found using the
classical action
\rescact\ evaluated on the classical solution $h(s,\rh,\ch)$
\eqn\twopt{S_b[h(s,L,\ch)]=Vm_s^6 L^3 {\p_{\rh}f(s,\rh,\ch)_{\rh=L} 
\over f(s,L, \ch)} }

We subtract from \twopt\ the action evaluated on the `free' solution
$\left({\rh\over L}\right)^{\b_+(s)}$, $S^{(0)}_b=Vm_s^6 L^3\b_+(s)$
and then retain only the dominant term as $L\r \infty$ (the toy model
of Appendix C clearly motivates this prescription)
\eqn\twoptt{\Pi^L(s)={L^2}[\b_-(s)-\b_+(s)]D(s)
L^{\b_-(s)-\b_+(s)} ={L^{-2\b_+(s)}}2\sqrt{1-s}D(s) .}
The renormalized correlation function is obtained by removing
the $L$ dependence, and is given by
\eqn\twopttr{\Pi(s)=2\sqrt{1-s}D(s).}
The renormalized correlation function thus defined
agrees with the reflection $S$ matrix element 
computed in the previous section. 
 
For small $s$ \twopttr\ becomes 
$$\Pi(s)=analytic +{ \pi^2\over 6}s^3\ln s + smaller \ \ terms.$$
This is exactly the two point function of the energy momentum tensor
of the M5 theory, which couples to a massless minimally coupled scalar
in $AdS_7$. This is a consistency check, since the IIA little string
theory reduces to the (0,2) theory at low energies.

\subsec{The implications of momentum dependent renormalizations} 

In order to obtain a cut off ($L$) independent $\Pi(s)$ from the bare
two point function $\Pi^L(s)$ we had to perform a momentum dependent
multiplicative renormalization. An additional finite momentum
dependent renormalization would change the formula \twopttr\ for
$\Pi(s)$. We have chosen our renormalization scheme to yield a two
point function that agrees with the $S$ matrix computed in section 3.
(Strictly, this might leave an $s$ dependent phase ambiguity.)

A momentum dependent renormalization 
is not multiplicative in position space. This may indicate the absence
of a natural definition of the correlation function in 
position space, hinting at nonlocality of the theory \refs{ \rpp, 
\rbanks} at scale ${m_s \over \sqrt{N}}$. If correct, this effect must
be distinct from the nonlocality at  $ m_s$ noted in
\rseiberga.

\newsec{Absorption from Infinity}

In this section we consider NS5-branes in IIA theory with small
but nonzero asymptotic string coupling $g_0$. The tube in the 
NS5-brane geometry is now of finite length, and the 
asymptotically flat part of the geometry cannot be ignored.  
We compute the absorption probability for a particle incident onto the 
NS5-brane from asymptotic infinity.

Consider a wave with $s>1$ incident (with zero momentum along the
branes and the sphere) onto the 5-brane. This wave may be partially
reflected at two locations - at the entrance to the tube from
asymptotic infinity, with a reflection amplitude $R(s)$, and at the
entrance to the AdS region from the tube, with a reflection amplitude
$D(s)$.

In order to be absorbed a wave must penetrate the tube, and then 
either be absorbed, or be reflected an even number of times and
then absorbed.
Thus the absorption amplitude is
$$A_{\infty}=\sqrt{(1-|D|^2)(1-|R|^2)}\sum_{n=0}^{\infty}
(e^{i\gamma}DR)^n.$$ 
$e^{i\gamma}=({\sqrt{N} \over g_0})^{2i\sqrt{s-1}}$ represents the
phase picked up by the wave traversing twice the length of the
tube. The absorption probability,
${\CF_{\infty}={(1-|D(s)|^2)(1-|R(s)|^2)\over
|1-e^{i\gamma}D(s)R(s)|^2}}$, is the ratio of flux absorbed by the
5-branes and the flux incident on them from asymptotic
infinity. $\CF_{\infty}$ may be computed by matching an exact solution
to the wave equation in the asymptotic and tube region with
\fgform. We find
\eqn\fll{\CF_{\infty}={(1-|D(s)|^2)(1-e^{-2\pi \sqrt{s-1}})\over
\big| 1+({
\sqrt{N} \over g_0})^{2i\sqrt{s-1}}D(s)e^{-\pi\sqrt{s-1}}
{\G(\b_{-}(s)+2) \over\G(\b_{+}(s)+2)} ({4\over s})^{i\sqrt{s-1}} 
\big|^2},} 
where $\G$ is the Gamma function. In particular this implies 
$|R(s)|=e^{-\pi\sqrt{s-1}}$, in agreement with \rklebd. 

The absorption cross section of the 5-branes is related to the absorption 
probability by 
$\s_{\infty}=4 \pi {\CF_{\infty} \over \o^3}$. 
Notice that, in contrast with a stack of M5-branes in flat space, the 
absorption cross section for a stack of NS5-branes is not zero in the 
limit $g_0 \r 0$ ($l_p \r 0$). This may be understood as follows. 
A particle incident on a  stack of isolated $M5$ branes in flat space has 
to tunnel through a broad potential barrier extending (in the usual 
coordinates) from $r={1\over \o}$ down to very small values of $r$. 
The suppression factor due to this barrier goes to infinity in the 
decoupling limit, so that, in order to send unit flux into the 
M5-branes one needs to shine an infinite amount of flux on
them from infinity. 

A stack of NS5-branes is, as we have seen, an array of stacks of
M5-branes.  The geometry very near any one element of the stack is
identical to that of an M5 in flat space, but is significantly
different for $r$ of order $R_{11}$. Since $R_{11} \r 0$ in the
decoupling limit, most of the potential barrier present in the
geometry of the isolated M5-brane is chopped off in this modified
geometry, and is replaced by the tube, through which particles with
$s>1$ propagate freely. The residual potential barrier results in only
finite tunneling suppression even in the decoupling limit. In order to
send unit flux down to the 5-branes, one needs to shine only a finite
amount of flux onto the branes from infinity.

For completeness we present also the small $g_0$ flux absorption ratio
$\CF_{\infty}(s)$ for $s<1$
\eqn\rat{\CF_{\infty}=({g_0\over \sqrt{N}})^{2\sqrt{1-s}}
{\pi^{4}\over 3}{|C(s)|^2 \over 2^{2\sqrt{1-s}}} {s^{3+\sqrt{1-s}}\over 
|\G(1+\sqrt{1-s})|^2}.}
The first factor in \rat\ is the tunneling suppression experienced by
the $s<1$ particle in penetrating through the tube of length ${g_0\over 
\sqrt{N}}$. Note that $\CF_{\infty} \r 0$ as $g_0 \r 0$. 

\bigskip

\cl{{\bf Acknowledgements}}
We would like to acknowledge useful discussions with O. Aharony,
M. Berkooz, T. Banks, C. Callan, O. Ganor, R. Gopakumar, I. Klebanov,
S. Lee, J. Maldacena, S. Mathur, G. Moore, M. Rangamani, A.
Vishwanath, E. Witten and
especially A. Strominger.  The work of S.M. was supported in part by
DOE grant DE-FG02-91ER40671 and N.S. by DOE grant DE-FG02-90ER40542.

\appendix{A}{Large $s$ Behavior}

In Minkowski space, for $u=\sqrt{\ch^2+\rh^2}\ll 1$ \mmeqgc\ describes
the propagation of a wave whose coordinate wavelength $\delta u$ is
approximately ${u^{{3\over 2}} \over \sqrt{s}}$.  The fractional
change in length of a wavelength over the length of one wavelength, of
order $ \sqrt{{u\over s}}$, is small for large $s$. Therefore, the WKB
approximation is valid for $s\gg 1$, and the dynamics of \mmeqgc\ is
that of null geodesics in \metrici.  

The geodesic equation for a particle moving in  \metrici\ with no 
velocity components
along the $S^3$ and the brane is
\eqn\geoa{\eqalign{
&Ay_0''-{1 \over 3}\p_{\rh}A\rh' y_0'-{1 \over 3}\p_{\ch}A \ch' y_0'=0 \cr
&A\rh''+{1\over 3}\p_{\rh}A(\rh')^2-{1\over 6A}\p_{\rh}A (y_0')^2
-{1\over 3}\p_{\rh}A (\ch')^2+{2\over 3}\p_{\ch}A\rh'\ch' =0\cr
&A\ch''+{1\over 3}\p_{\ch}A(\ch')^2-{1\over 6A}\p_{\ch}A (y_0')^2
-{1\over 3}\p_{\ch}A (\rh')^2+{2\over 3}\p_{\ch}A\rh'\ch' =0,\cr
}}
where $y_0$ is the time coordinate in \metrici\ and 
primes denote derivatives with respect to an affine parameter $\l$
along the curve.
A null geodesic also satisfies the mass shell condition
\eqn\nullgeo{y_0'=A\sqrt{\rh'^2+\ch'^2}.}  
In the tube $A={1\over \rh^2}$ and \geoa, \nullgeo\ admit the 
one parameter family of solutions 
\eqn\solutions{ \ch(\l)=\ch_0;  \  \ \rh(\l)=\l^{{3\over 2}}; \  \ 
y_0(\l)={3\over 2} \ln \l }  
that describe a light ray traveling down that tube at a fixed value of 
$\ch=\ch_0$, where $-\pi < \ch_0 \leq \pi$.  Numerical integration of
\geoa\ into the AdS part of the geometry indicates that 
every geodesic in \solutions\ except the one at $\ch_0=\pi$ reaches
$\ch=0, \rh=0$ (and so is absorbed by the M5-brane) at a finite value
of the affine parameter $\l_0$. As $\ch_0 \r \pi$, $\l_0 \r \infty$
and the geodesic at $\ch_0=\pi$ never reaches the horizon but is
reflected back.

Hence, classically, a light ray propagating down the tube 
with no momentum along the $S^3$ or the brane is absorbed by the 
NS5-brane with unit probability and $D\approx 0$.

\appendix{B}{Small $s$ Behavior}
We wish to determine $C(s)$ and $D(s)$ for small $s$. To do this we 
must find $f(s, \rh, \ch)$, the solution to \mmeqgc\ subject to the 
conditions listed in section 2.5. 

At $s=0$ \mmeqgc\ has two linearly independent solutions that, for 
small $\rh, \ch$ are functions only of $u^2=\rh^2+\ch^2$. They are
$$\psi_1=1, \qquad \psi_2=\sum_{n=-\infty}^{\infty}{\pi \over [\rh^2
+(\ch-2\pi n)^2]^{{3\over 2}}}.$$
For large $\rh$ $\psi_2={1\over \rh^2}$ and for small $u$
$\psi_2=\pi({1\over u^3}+{2\zeta(3)\over (2\pi)^3})$. 

$f(s=0, \rh, \ch)$ is a linear combination of $\psi_1$
and $\psi_2$.  Regularity at $u=0$ sets the coefficient of $\psi_2$ to
zero, and the normalization condition in sec 2.5 sets the coefficient of 
$\psi_1$ to unity. Therefore, $C(s=0)=1$ and $D(s=0)=0$.

We iterate our solution to successively higher order in $s$. Let 
$$f=f_0+f_1+f_2+...$$
$$C=C_0+C_1+C_2+...$$
$$D=D_0+D_1+D_2+...,$$
where $f_n, C_n, D_n$ are of successively of higher order in $s$ 
(later in this appendix we will find that $f_n, C_n, D_n$ are 
each of the form $s^nP_n(\ln s )$ where $P_n$ is a polynomial of 
degree $[{n\over 3}]$). Above we have 
found $f_0=\psi_1$, $C_0=1$, $D_0=0$. 
$f_{n+1}$ is obtained from $f_n$ by solving \mmeqgc\ iteratively, 
\eqn\iterat{-\left(\p_{\ch}^{2}+{1\over {\rh}^{3}}\p_{{\rh}}
{\rh}^{3}\p_{{\rh}}\right) f_{n+1}= 
\left(\sum_{m=-\infty }^{\infty}{s \pi \over [{\rh}^{2}+
(\ch-2\pi m)^2]^{{3\over 2}}}\right) f_{n}.}
\iterat\ does not uniquely determine $f_{n+1}$, since the differential 
operator on the LHS of \iterat\ has two zero modes, $\psi_1$ and
$\psi_2$. 
At large $\rh$ $f=\rh^{\b_+(s)}+D(s)\rh^{\b_-(s)}$ may be expanded as 
$$f= 1-s{\ln \rh \over 2}+{s^2\over 8}[ (\ln \rh )^2 -\ln \rh  ]+...
+{1\over \rh^2} (D_0+D_1+D_2+...)\{1+s{\ln \rh \over 2}+{s^2\over 8}
[ (\ln\rh )^2 +\ln \rh]+ ... \},$$
and so  the freedom to add an arbitrary multiple of $\psi_1$ to any 
solution of \iterat\ is fixed 
by the condition that the coefficient of the constant 
term in an expansion of $f_{n+1}(\rh)$ at large $\rh$ is zero. Let
$\t_{n+1}$ 
be a particular solution of \iterat\ obeying this asymptotic condition. 
Then
$$f_{m+1}=\t_{m+1}+B_{m+1}\psi_2,$$ 
where $B_{m+1}$ is a constant determined 
by imposing regularity of $f$ at small $u$.
Regularity of $f$ at $u=0$ does  
not imply regularity of $f_m$ for any $m \neq 0$.  It implies that
$f_m$ is proportional to the appropriate term in the small 
$s$ expansion of 
$C(s)\left({-s\pi\over u}\right)^{3\over 2}K_3\left(2 \sqrt{{-s\pi
\over u}} \right)$. 
Expanding this order by order in $s$ we find
\eqn\smallform{f(s,\rh,\ch)=(C_0+C_1+C_2...) \{ 1+{s\pi\over 2u} 
+{s^2\pi^2\over 4 u^2} -{s^3 \pi^3\over 12 u^3}[\ln({-s\pi\over u})-
\psi(1)-\psi(4)]+...\}.}
Matching with this form fixes $B_{m+1}$ and $C_{m+1}$ at each order. 

At first order we find
$$f_1=s\lim_{M\r \infty}\left[\left(\sum_{n=-M}^M{\pi \over 2 
\sqrt{\rh^2+(\ch-2\pi n)^2}}\right)-\half\ln (4\pi M )
\right].$$ 
At large $\rh$ $f_1=-{s\over 2}\ln \rh$. 
At small $\rh$ $f_1=s\{{\pi \over 2u}+\half[\g -\ln (4\pi) ]\}$ where $\g$ is 
Euler's constant. Therefore, 
$$C_1={s\over 2}[\g -\ln (4\pi)], \  \ D_1=0.$$
This procedure may be iterated to higher orders.

$ f_i, C_i, D_i \propto s^i$ for $i\leq 2$.  Since $f_2 \propto s^2$,
it is possible to choose $\t_3 \propto s^3$. However, at small $u$
$f_3 \approx s^3h(u) + { \pi^3\over 12 u^3}{s^3\ln s }$, and
therefore, we must choose $B_3=As^3+{\pi^2 \over 12}s^3\ln s $, where
$A$ is a constant.  This implies $D_3=(const)s^3+{ \pi^2\over
12}{s^3\ln s} $ and $C_3=(const)s^3+{\zeta(3) \over 48}s^3\ln s
$. Since at small $u$ $f_3 =s^3h(u) + {\pi^3 s^3\ln s \over 12 u^3}$,
it is possible to choose $\t_4=g(u)s^4+ s^4\ln s ({c_1\over u}+{c_2
\over u^4}+c_3)$ for small $u$. To ensure matching with \smallform\
each of $C_4, B_4$ must contain a term proportional to $ s^4\ln s $.
The situation is similar at the next order. However, at sixth order,
matching with \smallform\ (specifically the cross term from $C_3$ and
the ${1\over u^3}$ term) forces us to include also a term proportional
to $s^6(\ln s )^2$ in $B_6$, and hence in $C_6, D_6$.

Continuing successively to higher orders, it is clear that $D_i$ and
$C_i$ are of the form $s^i P_i(\ln s )$, where $P_i(x)$ is a
polynomial of order $[{i\over 3}]$ in $x$.  In other words, at small
$s$, $D(s)$ and $C(s)$ may each be written as a power series in $s$
and $s^3\ln s $.

\appendix{C}{A Toy Model}

In the string frame, the near horizon geometry of the NS5-brane is a 
semi-infinite tube, connected through an intermediate region to the 
M5-brane 
geometry. A quantum incident 
down the tube is either reflected back up the tube, or 
continues through the horizon of the 5-brane and disappears. 

In some ways this problem is similar to a free scalar field theory
with the action
\eqn\actions{S=\half \int dx dt\{(\p_0 \ph)^2-(\p_x\ph)^2
-\t(-x)\ph^2\} } 
in 1+1 dimensions. The region $x<0$ is analogous to the tube, $x>0$ is
similar to the M5 region, while the abrupt change in mass at $x=0$ is
the analogue of the transition region between the NS5 tube and the M5
region.

In this Appendix we explore this toy model. We compute the $S$ matrix
of this model in three different ways: using the LSZ formula,
computing flux ratios, and computing the Euclidean action as a
function of boundary values. We then consider a holographic projection
of this theory, and demonstrate that the two point function of the
boundary operator is equal to the reflection $S$ matrix of the bulk
particle.

\subsec{ The Propagator}

The Euclidean space propagator
\eqn\euclpro{G(x,y,s)=-\int dt e^{i(t-t')\sqrt{-s}}
\langle \phi(x,t)\phi(y,t')\rangle } 
is easily found by solving the appropriate
differential equation with the boundary condition that it vanishes as
either 
$|x|$ or $|y|$ go to infinity.  Our conventions are such that $s$ is
real and positive for Lorentzian energies and real and negative for
Euclidean energies.  We find
\eqn\propexpI{G(x,y,s)={\cases{
G^0(x-y,s,m=1) + e^{\sqrt{1-s}(x+y)} A(s)  & $x,y\leq 0$ \cr
e^{\sqrt{1-s}x-\sqrt{-s}y} B(s)  & $x\leq 0\ ;\ y\geq 0$ \cr
G^0(x,y,s,m=0) + e^{-\sqrt{-s}(x+y)} \tilde A(s) & $x,y\geq 0$.\cr}} }
where
\eqn\abtila{\eqalign{
&A(s)=-{1 \over 2\sqrt{1-s}}\left(1-2s-2\sqrt{-s}\sqrt{1-s}\right) \cr
&B(s)=\sqrt{-s} - \sqrt{1-s} \cr
&\tilde A(s)={1 \over
2\sqrt{-s}}\left(1-2s-2\sqrt{-s}\sqrt{1-s}\right) ,\cr}}
and
\eqn\freempro{G^0(x-y,s,m)={-1 \over 2\pi} \int _{-\infty}^\infty dp
{e^{ip(x-y)}\over -s+p^2 +m^2 } ={-e^{-\sqrt{m^2-s}|x-y|}\over
2  \sqrt{m^2-s}}}
is a free propagator.  $G$ can also be written as
\eqn\propexpII{\eqalign{ &G(x,y,s)=\cr
&{\cases{
G^0(x-y,s,m=1) +G^0(x-0,s,m=1)G^0(0-y,s,m=1) \Gamma_{LL}(s) & $x,y\leq
0 $\cr 
G^0(x-0,s,m=1)G^0(0-y,s,m=0)\Gamma_{LR}(s) & $x\leq 0\ ;\ y\geq 0$ \cr
G^0(x,y,s,m=0) +G^0(x-0,s,m=0)G^0(0-y,s,m=0) \Gamma_{RR}(s)  &
$x,y\geq 0$ \cr}}}}
in terms of
\eqn\results{\eqalign{
&\Gamma_{LL}= -2\left((1-2s)\sqrt{1-s}-2(1-s)\sqrt{-s}\right) \cr
&\Gamma_{LR}= -4\left((1-s)\sqrt{-s}+s\sqrt{1-s}\right) \cr
&\Gamma_{RR}= 2\left((1-2s)\sqrt{-s}+2s\sqrt{1-s}\right). \cr}}
Equations \propexpI-\results\ are valid by analytic continuation on
the complex $s$ plane with square root functions defined as follows.
For $s=Xe^{i\alpha} (0\leq\alpha<2\pi)$ we set
$\sqrt{-s}=-i\sqrt{s}=\sqrt{X}e^{i{\alpha-\pi\over 2}}$. Similarly for
$s-1=Xe^{i\alpha}$ we set
$\sqrt{1-s}=-i\sqrt{s-1}=\sqrt{X}e^{i{\alpha-\pi\over 2}}$.  In
particular, for $s$ infinitesimally above the real axis \propexpI\ is
the Minkowskian propagator $G_M$ defined by
$$iG_M(x,y,s)=\int e^{i\sqrt{s} x^0}\langle T\ph(x,x^0)\ph(y,0)
\rangle _{Minkowski}dx^0.$$

\subsec{S Matrix from the Propagator}

When $s>1$ we have two kinds of in states: particles with momentum
$p_{in}=\sqrt{s-1}>0$ coming from the left and particles with momentum
$k_{in}=-\sqrt{s}<0$ coming from the right.  We also have two kinds of
out states: particles with momentum $p_{out}=-\sqrt{s-1}<0$ going to
the left and particles with momentum $k_{out}=\sqrt{s}>0$ going to the
right. We normalize states covariantly so that 
$\langle k_{in}|k'_{in}\rangle =2\sqrt{s}\d(\sqrt{s_{in}}-\sqrt{s'_{in}})$, 
$\langle p_{in}|p'_{in}\rangle =2\sqrt{s-1}\d(\sqrt{s_{in}}-\sqrt{s'_{in}})$, 
and similarly for out states. 
The covariant $S$ matrix with this normalization of states is obtained by
amputating the external propagators from 
\propexpII. It is given by 
\eqn\smat{\left(\matrix{\langle p_{out}|p_{in}\rangle &\langle p_{out}
|k_{in}\rangle  \cr
\langle k_{out}|p_{in}\rangle &\langle k_{out}
|k_{in}\rangle  \cr}\right)
=\delta(\sqrt{s_{in}}-\sqrt{s_{out}})
 \left(\matrix{\G_{LL}(s)&\G_{LR}(s) \cr
		\G_{LR}(s)& \G_{RR}(s) \cr}\right).} 
With our normalization, completeness of in states implies 
\eqn\complete{\int_0^1 d\sqrt{s}{|k_{in}\rangle \langle k_{in}|
\over 2\sqrt{s}} +
\int_1^\infty d\sqrt{s}\left({|p_{in}\rangle 
\langle p_{in}|\over 2\sqrt{s-1}}
+{|k_{in}\rangle \langle k_{in}|\over 2\sqrt{s}} \right)=1.}
Inserting \complete\ into the the scalar product 
between two arbitrary in states we verify that the $S$ matrix is
unitary  
\eqn\unitarity{\left(\matrix{\G^*_{LL}(s)&\G^*_{LR}(s) \cr
		\G^*_{LR}(s)& \G^*_{RR}(s) \cr}\right)
\left(\matrix{ {1\over 2\sqrt{s-1} }&0 \cr
		0& {1\over 2\sqrt{s}}\cr}\right)
\left(\matrix{\G_{LL}(s)&\G_{LR}(s) \cr
		\G_{LR}(s)& \G_{RR}(s) \cr}\right) 
\left(\matrix{{1 \over 2\sqrt{s-1}}&0 \cr
		0& {1\over 2\sqrt{s}}\cr}\right)=1.}
For $s<1$ all particles incident from the right are reflected, and the 
appropriately normalized $S$ matrix is a pure phase.

\subsec{ $S$ matrix through flux ratios}

The probability for a particle incident from the left at energy
$\sqrt{s}$ to continue thorough to $x=\infty$, $\CF$, may be computed
very simply. A purely in-going wave function for $x>0$ is
$$\psi(x)= \cases{ e^{ipx}+E(s)e^{-ipx} & $x\leq 0$ \cr
F(s)e^{ikx} & $x\geq 0$,\cr }$$
with
$$F={2\sqrt{s-1}\over \sqrt{s}+\sqrt{s-1}} =1+E , \  \ E={-\sqrt{s}+
\sqrt{s-1} \over \sqrt{s}+\sqrt{s-1}}, $$
and hence
$$\CF=1-|E|^2 ={4\sqrt{s}\sqrt{s-1}\over (\sqrt{s}+\sqrt{s-1})^2}.$$
This result is equivalent to the $S$ matrix of the previous
section. For instance, given \smat, the amplitude for reflection is
${\G_{LL} \over 2\sqrt{s-1}} = E(s)$ (where we have accounted for
state normalizations and the delta function).

\subsec{ $S$ matrix from the Euclidean Action }

The scattering solution used in the flux computation of the previous
section may be analytically continued to the complex $s$ plane
\eqn\soln{\psi(x)=\cases{ e^{-\sqrt{1-s}x}+E(s)e^{\sqrt{1-s}x} &
$x\leq 0$ \cr
F(s)e^{-\sqrt{-s}x} &  $x \geq 0$, \cr } }
where
$$F(s)={2\sqrt{1-s}\over \sqrt{-s}+\sqrt{1-s}} ; \  \
E(s)={-\sqrt{-s}+
\sqrt{1-s} \over \sqrt{-s}+\sqrt{1-s}}. $$

For $s$ real and negative we compute the action for this 
solution on $-L<x<\infty$.
$$S={\p\psi \over \psi}_{x=-L}=\sqrt{1-s}{e^{-2\sqrt{1-s}L}
E(s)-1 \over  e^{-2\sqrt{1-s}L}E(s)+1}. $$       
In terms of the action $S_0$ of the similar solution
of a free massive theory $\psi=e^{-\sqrt{1-s}x}$
\eqn\actcor{ \lim_{L\r \infty} {(S-S_0)\over e^{-2\sqrt{1-s}L}}=
\G_{LL}(s).}
Thus after a subtraction and renormalization, the Euclidean action 
reproduces the reflection $S$ matrix. 

\subsec{Effective theory on the boundary}

Consider a distinct but related physical theory in which a 
free particle of unit mass on the real line interacts with an operator $O$
in a quantum mechanical system situated at $x=-L$ via the interaction 
action 
$$S_{int}=\int dt \ph(-L,t)O(t).$$
We will look for an $O$ such that the  
$\ph(x,t)$ Greens functions computed at $x,y<-L$ are identical to
\propexpII.  

The Euclidean space $\ph$  Greens function is given by
\eqn\proptaa{\eqalign{
&\tilde{G}(x,y,t)=-{\d \over \delta J(x,t)}{\d \over \delta J(y,0)}
\int \CD \ph e^{-S_f(\ph)+\int d\tau \ph(-L, \tau) 
O(\tau)+ \int d\tau d\chi J(\chi,\tau) \ph(\chi,\tau)}|_{J=0} \cr
&=G^0(x-y,t,m=1) -\int dt' d\tau 
G^0(x+L,t-\tau,m=1)G^0(-L-y,t',m=1) \langle O(\tau) O(t')\rangle }}
($S_f(\ph)$ is the free Euclidean action for the $\ph$ field) provided
all higher $n$ point functions of $O$ vanish. 
Setting $\tilde{G}(x,y,s)=\int dte^{i\sqrt{s} t}\tilde{G}(x,y,t)$ 
we find
$$\eqalign{\tilde{G}(x,y,s)&=G^0(x-y,s,m=1) \cr
&-G^0(x-0,s,m=1)G^0(0-y,s,m=1)
e^{2\sqrt{1-s}L}\int dte^{\sqrt{-s} t} \langle O(t)O(0)\rangle.}$$
We choose
\eqn\bq{\int dte^{\sqrt{-s} t} \langle
O(t)O(0)\rangle=-e^{-2\sqrt{1-s}L}\G_{LL}(s)=S_0-S.} 
This ensures $\tilde{G}(x,y,s)={G}(x,y,s)$, and the dynamics of
$\ph(x,t)$ for $x<-L$ in the new system is identical to the dynamics
of $\ph$ governed by \actions\ for $x<-L$.  It should be stressed that
there is no simple quantum mechanical system with such correlation
functions of $O$.

\appendix{D}{Geodesic Completion of the Brane Metric\foot{This 
appendix was worked out in collaboration with A. Strominger.}}

In this appendix we describe the geodesic completion of the 
near horizon metric of the NS5-brane. As a preliminary,
following \rtownsend, we describe the completion of the full
(not merely near horizon) geometry of M2, M5 and D3 branes, and 
demonstrate that the completion of the  geometry of 
several separated M5-branes is regular.

\subsec{Coincident extremal M2,M5 and D3 branes \rtownsend } 

The metric of a single wedge in the geometry of a set of 
coincident M2, M5 or D3 branes is 
\eqn\metric{ds^2=A ^{-{2\over p+1}} dx^2_{p+1}+A^{{2\over
d-2}}(dr^2+r^2d\Omega^2_{d-1})}
($0<r<\infty $), where $p$ is the spatial dimension of the brane, 
$d$ the spatial dimension 
of the transverse space, and $A=1+({\Lambda 
\over r})^{d-2}$. The Penrose diagram for this patch of the 
geometry is a diamond whose edges on the right are 
$\CI^{\pm}$ and those to the left are horizons at $t=\pm \infty$, at finite
affine distance.   

In terms of $\zeta=Kr^{{p+1\over d-2}}$ (where $K$ is a constant)
the metric of the near horizon region $\zeta^{p+1} \ll \Lambda^{d-2}$
of \metric\ is
\eqn\metricb{ds^2=\Lambda^2 \left({p+1\over d-2}\right)^2\left(
\zeta^2 dx_{p+1}^2+ {d\zeta^2\over \zeta^2} \right)
+\Lambda^2 d\Omega_{d-1}^2.}
\metricb\ may be smoothly
continued past its horizons by extending the range of $\zeta$ to 
negative values, hence the same is true of \metric\ 
(written in terms of $\zeta$) for $|\zeta|\ll K\Lambda^{{p+1\over d-2}}$ . 
For larger $|\zeta|$ the
Harmonic function $A=1+{\Lambda^{d-2} \over
\zeta^{p+1}}$ behaves differently for odd and even $p$. 
$A$ is well behaved for all $\zeta$ when $p$ is odd, but vanishes 
at $\zeta= -\Lambda^{{d-2\over p+1}}$ when $p$ 
is even, leading to a curvature singularity in \metric\ at that point.

The Penrose diagram of \metric\ is a diamond. Depending on whether $p$
is even or odd our extension of the geometry augments the diamond
differently.  For $p$ odd we add another diamond whose bottom right
edge is attached to the top left edge of the original diamond. The two
left edges of the new diamond are new $\CI^{\pm}$. The upper right
edge is a new horizon at finite affine distance.  For $p$ even we add
a triangle, whose bottom right edge is attached to the top left edge
of the original diamond. The top right edge is a new horizon at finite
affine distance. The vertical line is the singularity.  In both cases
the extension to the original geometry has its own horizon at finite
affine distance, which may in turn be continued through.  The
corresponding Penrose diagrams are depicted in fig.\ 2.  Each point on
the diagram represents $S^{d-1} \times R^p$.

\fig{Penrose diagram of the geodesic completion of the metric of the
M5 and D3 brane (2a), and the M2 brane (2b). (2a) is also the Penrose
diagram for a $\ch=0$ slice of the near horizon region of the
NS5-brane.}  {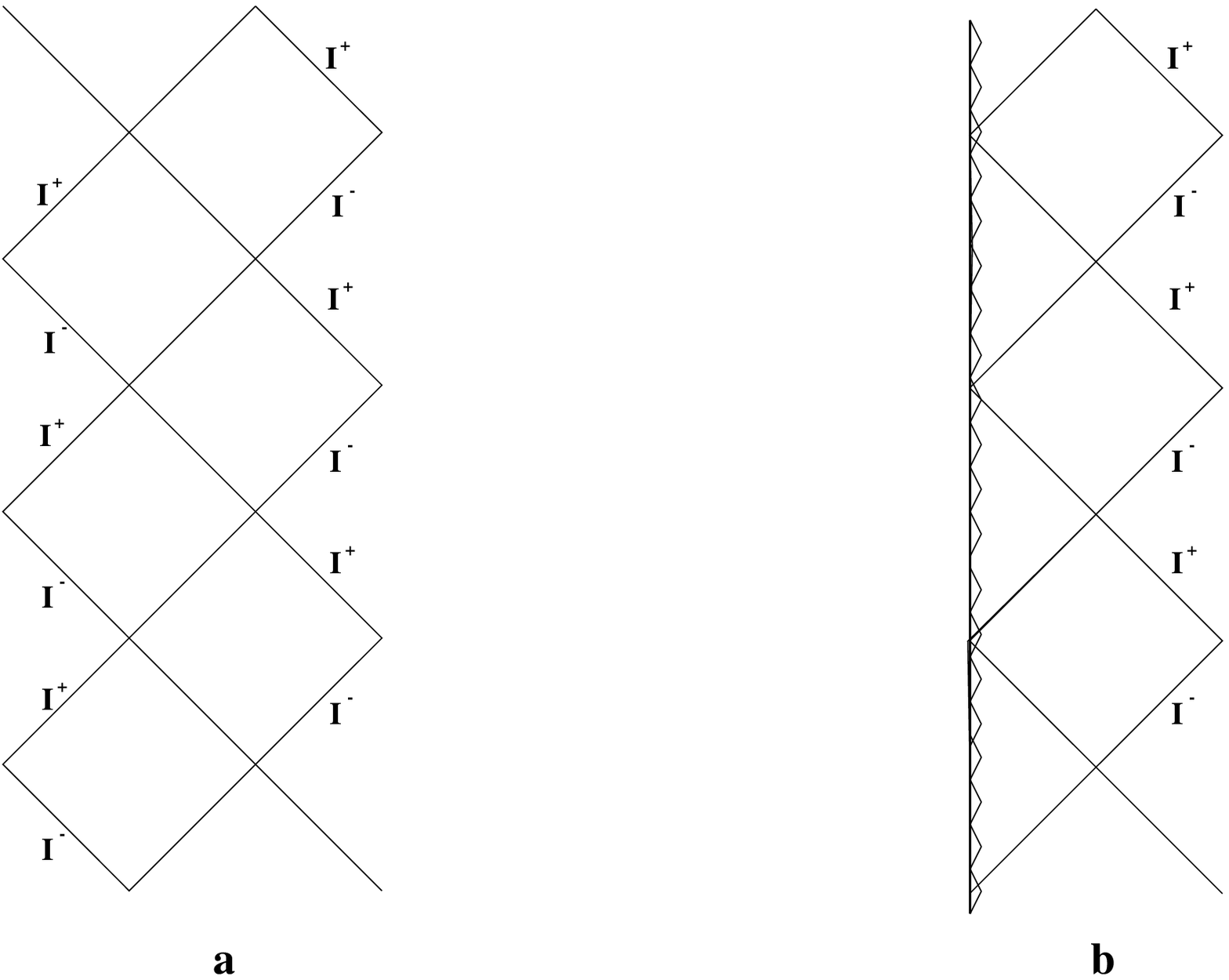} {4truein}

\subsec{Multiple M5 geometry \rtownsend } 

The metric corresponding to two non-coincident $p$ branes is
of the form  \metric.
The  harmonic function $A$ however picks up an additional term 
corresponding to the second brane. Let the first brane be located 
at the origin, and the second brane on the $z$ axis at $z=a$ 
in a Cartesian coordinate system in transverse space. 
The term in $A$ corresponding to 
the second brane is (recall $r=K\zeta ^{{p+1\over d-2}}$) 
\eqn\harmonic{A={\Lambda^{d-2}\over \left( K^2\zeta^{{2(p+1)\over d-2}} - 
2Ka \cos \theta \zeta^{{p+1\over d-2}} +a^2\right)^{d-2\over 2}},
\qquad \sin\t={z\over r}.}
We attempt to extend through the horizon of the first brane by
extending the range of $\zeta$ to negative values. This is permissible
only if ${p+1\over d-2}$ is an integer (in order for \harmonic\ to be
real).  If ${p+1\over d-2}$ is moreover an even integer, as is the
case for M5-branes, then the `mirror universe' ($\zeta < 0$) is
identical to the original in every respect. Therefore, this is also
true of any further extension through any other brane. The full space
is completely regular.

\subsec{The IIA NS5-brane}

Consider an array of M5-branes periodically identified.  
The identification is
regular in each wedge as every wedges has the geometry of \metricg. 
Since wedges in the multiple M5 geometry may be patched together in 
a regular manner we conclude that the full geometry is regular.

We focus on the near horizon region of the NS5-brane geometry. 
The geodesic completion of a $\ch=0$ slice of \metrici\ has the 
Penrose diagram depicted in fig.\ 2a, where $\CI^{\pm}$ represent  
light-like asymptotic infinity in the NS5-brane tube. 
As demonstrated in appendix A, null 
geodesics starting out in the tube at arbitrary values of $\ch$ are
qualitatively similar to those starting at $\ch=0$, and so fig.\ 2a
provides a fair picture of the causal structure of the spacetime.
Each point in the diagram may roughly be thought of as the product of 
$R^5$, an $S^3$ and an $S^1$. The $S^1$ shrinks to zero size at the 
boundaries of the diagram. 

\listrefs
\end